\documentclass[prl,aps,reprint,nopacs]{revtex4-1}
\usepackage{amsmath}
\usepackage{amsfonts}
\usepackage{amssymb}
\usepackage{revsymb}
\usepackage{graphicx}

\def\ket#1{|\,#1 \,\rangle}
\def\bra#1{\langle \, #1 \,|}

\begin{document}
\title{Developing a field independent frequency reference by averaging over transitions to multiple hyperfine levels.}
\author{M. D. Barrett}
\affiliation{Center for Quantum Technologies, 3 Science Drive 2, Singapore, 117543}
\affiliation{Department of Physics, National University of Singapore, 2 Science Drive 3, Singapore, 117551}
\email{phybmd@nus.edu.sg}
\begin{abstract}
We show that by averaging over transitions to multiple hyperfine levels, quadrupole shifts and dominant Zeeman effects exactly cancel whenever the nuclear spin, $I$, is at least as large as the total electronic angular momentum, $J$.  The average frequency thus defines a frequency reference which is inherently independent of external magnetic fields and electric field gradients.  We use $\mathrm{Lu}^+$ to illustrate the method although the approach could be readily adapted to other atomic species.  This approach practically eliminates the quadrupole and Zeeman shift considerations for many clock candidates.
\end{abstract}
\maketitle
The realisation of accurate, stable frequency references have enabled important advances in science and technology.  Well-known examples include the Global Positioning System, geodesy, and test of fundamental physical theories.  Increasing levels of accuracy continue to be made with atomic clocks based on optical transitions in isolated atoms \cite{AlIon,SrYe,HgIon,SrIon,YbIon,InIon,YbForbidden,RMP}.  By now a number of groups have demonstrated superior performance over the current caesium frequency standard with the best clocks to date having inaccuracy at the $10^{-18}$ level \cite{AlIon,SrYe}.  To date, all optical clocks are based on the frequency of a single atomic transition. For this reason transitions between $J=0$ levels have had a prominent role in the development of optical clocks due to their inherent insensitivity to electromagnetic fields.  In almost all cases averaging over near-degenerate Zeeman transitions is used to cancel or deduce residual shifts from external fields.  This need not be the only approach: averaging over multiple, non-degenerate transitions can provide a frequency standard less susceptible to external perturbations.

Recently it was shown that averaging over two clock transitions could be used to suppress blackbody radiation (BBR) shifts \cite{BBRShift}.  The basic idea of the method relies upon the fact that a frequency comb, when referenced to two different frequencies $f_1$ and $f_2$, provides access to frequencies of the form $f_s=f_1+\frac{n}{m}f_2$ for integers $n,m$.  When $f_1$ and $f_2$ vary due to an external influence, a suitably chosen pair of integers provides a frequency $f_s$ in which the external influence can be substantially minimised.  In the spirit of this approach, we show that averaging over transitions to multiple hyperfine levels can lead to an exact cancellation of quadruple shifts and dominant Zeeman effects whenever $I\geq J$, where $I$ is the nuclear spin and $J$ is the total electronic angular momentum. This practically eliminates Zeeman and quadrupole shifts in developing a frequency standard by realising an effective $J=0$ level.  We start with a description of the general idea, treating first the Zeeman shifts from external magnetic fields and then the quadrupole shifts from electric field gradients. We then illustrate our method using $\mathrm{Lu}^+$ as a concrete example but the method would be applicable to many other atomic species.

Let us first consider the effects of an external magnetic field.  The Hamiltonian for a given fine structure level is block diagonal in $m_F$. Each block has the general form
\begin{equation}
H=H_0 + H_1 + H_2
\end{equation}
where $H_0$ is a diagonal matrix of the zero field energies $E_F$, $H_1$ is diagonal with entries $g_F m_F \mu_B B$ corresponding to the first order Zeeman shifts, and $H_2$ has only off-diagonal entries responsible for deviations from the from the first order Zeeman effect.  If we consider the average value of the diagonal elements, $H_2$ does not contribute  and we are left with 
\begin{equation}
\langle H_{F,F} \rangle = \langle E_F \rangle+m_F \langle g_F \rangle \mu_B B
\end{equation}
If $I>J$, it is easily shown that $\langle g_F \rangle = g_I$.  This average is an invariant of $H$ and hence, for a given $|m_F|\leq I-J$, the average shift over all $F$ states is simply $m_F g_I \mu_B B$ identical to that for a $J=0$ level.  When $I=J$ the result still applies and we are restricted to the consideration of $m_F=0$ states.  

The quadrupole shift due to electric field gradients can be treated as a perturbation on the previous result, as the shift is typically much less than the Zeeman splitting between $m_F$ levels. The shift for each $\ket{F,m_F}$ state is simply the diagonal elements of the interaction given, in the low field limit, by \cite{ItanoQuad} 
\begin{widetext}
\begin{multline}
\bra{F,m_F}H_Q\ket{F,m_F} = (-1)^{2F-m_F+I+J} (2F+1) \begin{pmatrix}F & 2 & F\\ -m_F & 0 & m_F\end{pmatrix} \begin{Bmatrix}F & 2 & F \\ J & I & J\end{Bmatrix}\\
\times \begin{pmatrix}J & 2 & J\\ -J & 0 & J\end{pmatrix}^{-1}\Theta(J) \left[ A(3\cos^2\beta-1+\epsilon \sin^2\beta (\cos^2\alpha-\sin^2\alpha))\right].
\end{multline}
\end{widetext}
In this expression, the first line determines the relative size of the shift for each of the $\ket{F,m_F}$ states.  The terms in square parentheses on the second line concerns only the trap geometry with respect to the quantisation axis.  Specifically, the electric potential, $\Phi$, in a neighbourhood of the atom, is given in the principal-axis coordinates by
\begin{equation}
\Phi(x,y,z)=A[x^2+y^2-2 z^2+\epsilon(x^2-y^2)].
\end{equation}
The Euler angles, $\alpha$ and $\beta$, determine the rotation of the principal-axis coordinate system with respect to the laboratory frame defined by the quantisation axis.  The rest of the terms on the second line characterise the magnitude of the quadrupole coupling with $\Theta(J)$ giving the quadrupole moment for the rem of interest as defined in \cite{ItanoQuad}.   

It is known that averaging the quadrupole shift over all $m_F$ states for a fixed $F$ yields zero \cite{ItanoQuad}. This follows immediately from the expression above and well-known $3j$-symbol identities.  It has also been pointed out that averaging over three orthogonal spatial orientations of the quantisation axis also gives zero \cite{ItanoQuad}.  However this approach is limited by the accuracy at which the the field orientations can be set \cite{HgIon}.  What is perhaps less well known is that averaging over all hyperfine states for a fixed $m_F$ also gives zero provided $I \geq J$ and $|m_F|\leq I-J$.  Under these conditions, the expansion of $\ket{F,m_F}$ in the $IJ$ basis includes all possible values of $m_J$ and may be written
\begin{equation}
\label{basis}
\ket{F,m_F}=\sum_{m_J} C_{F,m_J} \ket{I,J,m_F-m_J,m_J}.
\end{equation}
From this expansion we have
\begin{equation}
\label{HQtrace}
\begin{split}
&\sum_F \bra{F,m_F}H_Q\ket{F,m_F}\\
&=\!\sum_{\substack{F\\m_J,m'_J}}\!\!C_{F,m_J} C_{F,m'_J} \bra{J,m'_J}H_Q\ket{J,m_J}\delta_{m'_J,m_J}\\
&=\sum_{F,m_J} C^2_{F,m_J} \bra{J,m_J}H_Q\ket{J,m_J}\\
&=\sum_{m_J} \bra{J,m_J}H_Q\ket{J,m_J}.
\end{split}
\end{equation}
where we have used that fact that $H_Q$ is independent of nuclear spin and $\sum_F C^2_{F,m_J}=1$.  Since the average of the quadrupole shift over all $m_J$ is zero \cite{SrIon, ItanoQuad}, it therefore follows that the shift vanishes when averaged over all $F$.  This result holds even when there is significant Zeeman mixing of the hyperfine states.  The $F$ in Eq.~\ref{HQtrace} is then simply a label for the $2J+1$ eigenstates associated with the particular $m_F$.  We also note that the average over all $m_J$ for any tensor operator is zero, which follows directly from the Wigner-Ekart theorem.  Hence Eq.~\ref{HQtrace} is quite generally applicable. 

The averaging we have described leads to an effective $J=0$ level in so far as magnetic fields and electric field gradients are concerned.  This provides a greater degree of flexibility in considering potential clock transitions. It allows one to capitalise on other favourable properties of a transition and still retain the benefits provided by a $J=0$ level.  We now illustrate these considerations using Lu$^+$ as a concrete example.

\begin{figure}
\includegraphics{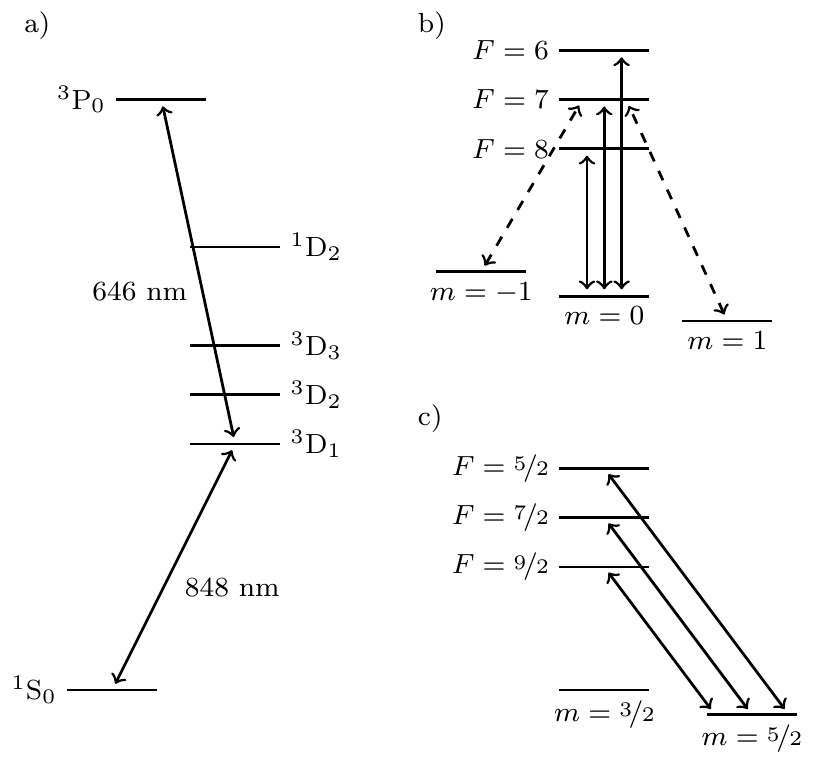}
\caption{a) The level structure of Lu$^+$.  The $^3$D$_1$ to $^3$P$_0$ transition at $646\,\mathrm{nm}$ is a nominally cycling transition providing state detection, cooling and state preparation.  The $^1$S$_0$ to $^3$D$_1$ at $848\,\mathrm{nm}$ provides a high $Q$ clock transition ($Q\sim 2.4 \times 10^{20}$). b)  Averaging over the three transitions $\ket{^1\mathrm{S}_0,7,0}$ to $\ket{^3\mathrm{D}_1,F',0}$ in $^{176}\mathrm{Lu}^+$ indicated by solid arrows realises an effective $J=0$ to $J=0$ transition as described in the text.  The $M1$ forbidden transition $\ket{^1\mathrm{S}_0,7,0}$ to $\ket{^3\mathrm{D}_1,7,0}$ is realised by averaging over the $\ket{^1\mathrm{S}_0,7,\pm 1}$ to $\ket{^3\mathrm{D}_1,7,0}$ transitions indicated by dotted arrows. c) The average over the three transitions $\ket{^1\mathrm{S}_0,7/2,5/2}$ to $\ket{^3\mathrm{D}_1,F',3/2}$ in $^{175}\mathrm{Lu}^+$ is field independent at approximately $4750\,\mathrm{G}$.}
\label{LuStructure}
\end{figure}

We have recently begun to explore singly ionised Lutetium as a possible clock candidate.  The ion has a similar level structure to neutral Barium with a spin singlet ($^1\mathrm{S}_0$) ground state and a low lying triplet of $D$ levels as illustrated in Fig.~\ref{LuStructure}.  Our focus is on the highly forbidden $M_1$ transition, $^1\mathrm{S}_0$ to $^3\mathrm{D}_1$, at approximately $848\,\mathrm{nm}$. Recent calculations indicate a lifetime of approximately $23$ hours giving a $Q$-value of $2.4 \times 10^{20}$ \cite{Calc1}.  The differential static polarisability, $\Delta\alpha$, gives a fractional blackbody radiation shift of $5.4\times10^{-17}$ at room temperature and the sign of $\Delta\alpha$ allows for the possibility of eliminating micro-motion effects with the appropriate choice of trap drive frequency \cite{Micromotion, SrIon}.  The $^3\mathrm{D}_1$ to $^3\mathrm{P}_0$ transition has a linewidth of $2\pi\times 2.45\,\mathrm{MHz}$ which allows for both clock state detection and a low Doppler cooling limit relative to most ion transitions.  

The clock transition, with the $J=1$ excited state, would be limited by its interaction with magnetic fields and electric field gradients \footnote{Calculations reported in \cite{Calc1} give a value of zero. However we believe this is most likely in error as there appears no fundamental reason for it to vanish exactly.}.  However, from the previous arguments, the average value of the frequencies, $\nu_{F'}$, corresponding to the transitions $\ket{^1\mathrm{S}_0,F,m_F}$ to $\ket{^3\mathrm{D}_1,F',m_F}$ would be free of these effects.  This, of course, neglects coupling to other fine structure levels which results in a residual quadratic shift.  This is due almost entirely to the $^3\mathrm{D}_2$ level and is given by
\begin{equation}
h\delta\nu= -\frac{\left[\mu_B B(g_L-g_S)\right]^2}{2 \hbar \omega_{FS}},
\end{equation}
where $\omega_{FS}=2\pi\times19.16\,\mathrm{THz}$ is the fine-structure splitting between the $^3\mathrm{D}_1$ and $^3\mathrm{D}_2$ levels.  Taking $g_L\approx1$ and $g_S\approx 2$ we find a residual quadratic shift of just $50\,\mathrm{mHz/G^2}$.  What remains to consider is the field dependence of the individual component transitions which affects the short term stability of the average frequency.

Lutetium has two naturally occurring isotopes,  $^{175}\mathrm{Lu}$  and $^{176}\mathrm{Lu}$, with nuclear spins $I=7/2$ and $I=7$ respectively. The latter provides the possibility of using $m_F=0$ transitions, as shown in Fig.~\ref{LuStructure}(b) which are inherently field insensitive at low field.  With its large hyperfine splittings of approximately $10\,\mathrm{GHz}$ \cite{LuHyperfine}, the quadratic Zeeman shifts of the $\ket{F',0}$ states are approximately $23.2, -1.5,$ and $-21.7\,\mathrm{Hz/G^2}$  for $F'=6,7$ and $8$ respectively.  At an operating field of $0.1\,\mathrm{G}$ the component transitions  $\ket{F,0}\rightarrow\ket{F',0}$ have field sensitivities below $5\,\mathrm{Hz/G}$.  However it should be noted that the $\ket{F=7,0}\rightarrow\ket{F'=7,0}$ transition is M1 forbidden but it can be realised as an effective transition by averaging $\ket{F=7,m_F=\pm1}\rightarrow\ket{F'=7,0}$.  This gives a faithful representation of the forbidden transition since the ground states do not contribute to either the quadratic Zeeman shift or the quadrupole shift. However the linear Zeeman shifts in the $^1\mathrm{S}_0$ states increase the field sensitivities to $g_I \mu_B/h \approx 350\,\mathrm{Hz/G}$ \footnote{We have used the value of $g_I=-2.46\times 10^4$.  The magnitude is taken from \cite{gIvalue} and the sign deduced from \cite{gIvalue2}.  We use the sign convention given in \cite{signConvention}}.   

Since the cancelling of both Zeeman and quadrupole shifts does not depend on the level of hyperfine mixing, one could consider defining the frequency reference at a well-defined, measurable, field independent point.  Of course careful consideration to the magnetic field sensitivity of the component transitions should be given.  Since the hyperfine splitting for Lu$^+$ is large, it is unlikely one could do better than with the $m_F=0$ states of $^{176}\mathrm{Lu}^+$ at low field.  However for the purposes of illustration we summarise considerations applicable to $^{175}\mathrm{Lu}^+$.

\begin{figure}
\includegraphics[width=8.6cm]{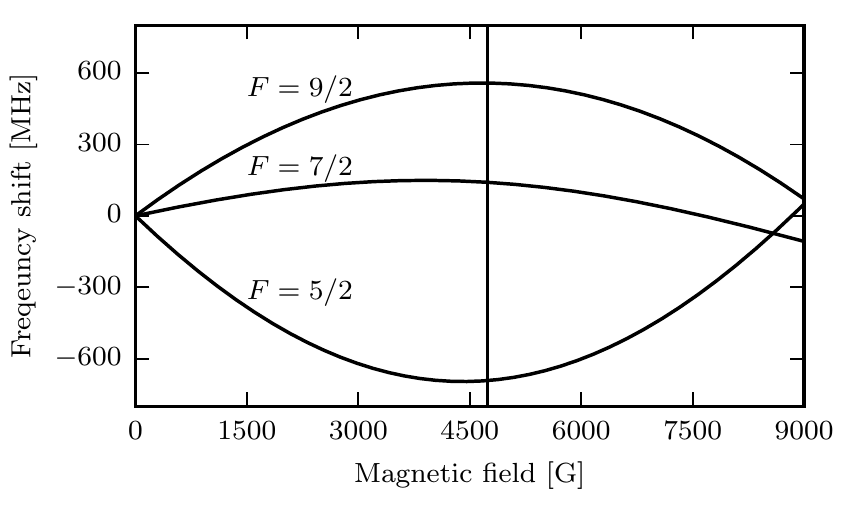}
\caption{Field dependence of the three transitions $\ket{^1\mathrm{S}_0,7/2,5/2}\rightarrow\ket{^3\mathrm{D}_1,F,3/2}$ relative to the zero field values.  The energies are labeled by the zero-field hyperfine quantum number, $F$. The vertical line indicates where the average of the three transitions becomes field independent.}
\label{FieldDependence}
\end{figure}

The isotope, $^{175}\mathrm{Lu}^+$, with nuclear spin $I=7/2$, has no $m_F=0$ states.  Hence any single transition has a significant linear Zeeman shift which may compromise the ability to accurately achieve the desired average.   For the $\ket{^3\mathrm{D}_1,F,m_F\!=\!3/2}$ states the linear Zeeman shifts are approximately $-300, 66.7,$ and $233.3\,\mathrm{kHz/G}$ for $F=5/2, 7/2,$ and $9/2$ respectively.  However, if one were to use $\Delta m_F\neq 0$ transitions as shown in Fig.~\ref{LuStructure}(c), the average frequency has a field dependence of
\begin{equation}
\label{FIE}
h\delta\nu=g_I \Delta m \mu_B B-\frac{\left[\mu_B B(g_L-g_S)\right]^2}{2 \hbar \omega_{FS}}
\end{equation}
which is field independent at a field of approximately $4750\,\mathrm{G}$ \footnote{We have used the value of $g_I=-3.47\times 10^4$.  The magnitude is taken from \cite{gIvalue} and the sign deduced from \cite{gIvalue2}.  We use the sign convention given in \cite{signConvention}}.  In Fig.~\ref{FieldDependence}(c) we plot the frequency shifts of the three components transitions with the vertical line indicating the field independent point for the average frequency.  At this point, the linear shifts of the component transitions are reduced to $25.5,-20.3,$ and $-5.2\,\mathrm{kHz/G}$.  These sensitivities place stringent requirements on the magnetic field stability needed to determine the average. However, the most sensitive level is within a factor of 6 of the component transitions used in the Al$^+$ clock.  As noted in \cite{AlIon2} this impacts on short term stability but not the long term inaccuracy.

The approach we have suggested here applies when $I \geq J$ and this is true for the ground state manifold as well provided the appropriate averaging can be done.  Consider for example $^{88}\mathrm{Sr}^+$ which has a $S_{1/2}$ to $D_{5/2}$ clock transition.  This ion requires averaging over six different Zeeman transitions in order to cancel magnetic field and quadrupole shifts, which places stringent requirements on magnetic field stability \cite{SrIon,SrIonLock}.  Alternatively $^{87}\mathrm{Sr}^+$ has $m=0$ states that would be inherently more magnetically stable. This ion has two ground states with $F=4,5$ and six excited states with $F=2,...,7$.  The average of the six transitions $\ket{4,0}$ to $\ket{F=2k,0}$ and $\ket{5,0}$ to $\ket{F=2k+1,0}$ for $k=1,2,3$ provides exactly the average needed to eliminate the magnetic field effects in both ground and excited states, and the quadrupole shifts of the D state.  This approach uses the same amount of averaging as for $^{88}\mathrm{Sr}^+$, but would presumably be more stable.

It is of interest to note that the requirement $I\geq J$ can be seen as a generalisation of the requirement $I\neq 0$ for the $J=0$ cases such as Al$^+$ and neutral strontium.  The requirement $I\geq J$ results in an average over $2J+1$ states to obtain an effective $J=0$ level.  In so far as the the average frequency is concerned, any $J$ level can be treated on an equal footing to that of a $J=0$ level. In general, a field independent point can be found for the frequency reference defined by the average, with the $J=0$ case requiring just one transition.  The field independent point is governed by an equation of the form given in Eq.~\ref{FIE}, with only a slight modification needed for the $J=0$ case to account for hyperfine induced changes in the $g$-factor.  Notably, the curvature at the field independent point is governed solely by the fine-structure splitting and is independent of the nuclear spin.  Hence there is no significant advantage to using integer spin for the $J=0$ case. For $J\neq 0$, one needs to consider the field dependence of the component transitions used to obtain the average which impacts on the short term stability.  Earlier proposals \cite{TlIon, AlIon3} aiming for the integer spin candidates were based on the notion of using $m=0$ to $m'=0$ transitions at low fields.  These transitions are fundamentally forbidden and require a magnetic field to induce a non-zero coupling \cite{YbForbidden2}.  At a fundamental level one could simply go to a field independent point and obtain the same result.  The only exceptional case would be $I=J=0$ for which the argument does not hold \cite{YbForbidden,YbForbidden2}.

In conclusion, we have shown that averaging over transitions to multiple hyperfine levels can lead to an exact cancellation of quadruple shifts and dominant Zeeman effects when $I\geq J$. Such averaging provides an effective $J=0$ level and therefore a more practical approach to the cancellation of important shifts of clock frequencies.  For the case of Lutetium averaging over $m_F$ states to cancel the quadrupole shift would be complicated by magnetic field considerations and involve no less than 11 transitions.  Averaging over just three transitions, we cancel both quadrupole and magnetic field shifts.  As with any other clock which uses averaging over multiple Zeeman states, we would not expect clock stability to be adversely affected by the averaging. For completeness we also note that this approach could be extended across multiple fine structure levels although such averaging would typically require a frequency comb.
\begin{acknowledgements}
We would like to thank Markus Baden for help with preparation of the manuscript. We also thank David Wineland for helpful discussions.  We acknowledge the support of this work by the National Research Foundation and the Ministry of Education of Singapore.
\end{acknowledgements}

\end{document}